\documentclass[12pt,a4paper,titlepage=false]{scrartcl}

\usepackage{amsmath,amssymb,url}
\usepackage{graphicx} 
\usepackage{booktabs}
\usepackage{array}
\usepackage{caption}
\usepackage{rotating}
\usepackage{xspace}
\usepackage{units}
\usepackage{xcolor}
\usepackage{slashed}
\usepackage{acronym}
\usepackage{enumitem}     
\usepackage{bookmark}
\usepackage{siunitx}
\usepackage[
  font=small,
  labelfont=bf,
  format=plain,
  margin=1em,
  indention=1em,
  labelsep=space,
  skip=5em]{caption}
\usepackage{hyperref}
\usepackage{cleveref}
\usepackage{multirow}
\hypersetup{%
    colorlinks,
    citecolor=blue,
    filecolor=black,
    linkcolor=black,
    urlcolor=black
}

\usepackage{cleveref}
\usepackage[titletoc]{appendix}
\def\email#1#2{\thanks{\texttt{#1@{}#2}}}

\newcommand\simge{\mathrel{%
   \rlap{\raise 0.511ex \hbox{$>$}}{\lower 0.511ex \hbox{$\sim$}}}}
\newcommand\simle{\mathrel{
   \rlap{\raise 0.511ex \hbox{$<$}}{\lower 0.511ex \hbox{$\sim$}}}}

\newcommand\be{\begin{equation}}
\newcommand\ee{\end{equation}}
\newcommand\bea{\begin{eqnarray}}
\newcommand\eea{\end{eqnarray}}
\newcommand\ba{\begin{array}}
\newcommand\ea{\end{array}}

\newacro{NWA}{narrow-width-approximation}
\newacro{FtW}{finite-top-width}
\newacro{NLL}{next-to-leading-logarithmic}

\newcommand\sqrts{\ensuremath{\sqrt{s}}\xspace}

\newcommand{\GeV}{{\ensuremath\rm GeV}}

\newcommand{\ValGeV}[1]{\unit[#1]{GeV}}

\newcommand\epem{e^+e^-}
\newcommand\wbwb{\ensuremath{b\, W^+\, \bar{b}\, W^-}\xspace}
\newcommand{\Hwbwb}{\ensuremath{\wbwb H}\xspace}

\newcommand\ttb{\ensuremath{t\,\bar{t}}\xspace}
\newcommand{\tth}{\ensuremath{t\,\bar{t}\,H}\xspace}

\newcommand{\eett}{\ensuremath{\epem \to \ttb}\xspace}

\newcommand{\eewwbb}{\ensuremath{\epem \to \wbwb}\xspace}
\newcommand{\eewwbbH}{\ensuremath{\epem \to \Hwbwb}\xspace}

\newcommand{\wz}{\textsc{Whizard}\xspace}

\newcommand{\po}{\textsc{Powheg}\xspace}
\newcommand{\om}{\textsc{O'Mega}\xspace}
\newcommand{\ol}{\textsc{OpenLoops}\xspace}

\newcommand{\re}{\textsc{RECOLA}\xspace}

\newcommand{\go}{\textsc{GoSam}\xspace}

\newcommand{\hepmc}{\texttt{hepmc}\xspace}
\newcommand{\rivet}{\texttt{Rivet}\xspace}
\newcommand{\toppik}{\texttt{Toppik}\xspace}

\newcommand{\ket}[1]{\ensuremath{\left|{#1}\right\rangle}}
\newcommand{\bra}[1]{\ensuremath{\left\langle{#1}\right|}}
\newcommand{\braketop}[3]
  {\ensuremath{\bra{#1\vphantom{#2#3}\,}#2\ket{#3\vphantom{#1#2}}}}
\newcommand{\T}[1]{\mathrm{#1}}
\newcommand{\MEprod}{\ensuremath{\mathcal{M}_\T{prod}}\xspace}
\newcommand{\ME}{\ensuremath{\mathcal{M}}\xspace}

\newcommand{\abs}[1]{\ensuremath{\left\lvert#1\right\rvert}}
\newcommand{\XSexpanded}{\ensuremath{\sigma_{\T{NRQCD}}^{\T{expanded}}}\xspace}
\newcommand{\XSmatched}{\ensuremath{\sigma_\T{matched}}\xspace}

\newcommand{\XSfo}{\ensuremath{\sigma_\T{FO}}}
\newcommand{\XSresummed}{\ensuremath{\sigma_\T{NRQCD}^\T{full}}\xspace}
\newcommand\no{\notag{}\\}

\newcommand{\AShard}{\ensuremath{\alpha_{\T{H}}}\xspace} 
\newcommand{\ASsoft}{\ensuremath{\alpha_{\T{S}}}\xspace}
\newcommand{\ASusoft}{\ensuremath{\alpha_{\T{US}}}\xspace}
\newcommand{\of}[1]{\left[{#1}\right]}
\newcommand{\switch}{f_s}

\newcommand{\XSNLO}{\ensuremath{\sigma_\T{NLO}}\xspace}

\newcommand{\XSNLONLL}{\ensuremath{\sigma_\T{NLO+NLL}}\xspace}
\newcommand{\realOfDiagrams}[3][]{
  \left(#1\diagram{#2}\right.\left\lmoustache\diagram{#3}\right)
}

\newcommand{\FFNLL}{\ensuremath{F_\T{NLL}}\xspace}

\newcommand{\FFNLLmO}{\tilde{F}_\T{NLL}}

\newcommand{\FFexpNLLmOne}{\tilde{F}^\T{exp}_\T{NLL}}
\newcommand\extraspace{\hspace{+1.3em}}
\newcommand\lessspace{\hspace{-1.3em}}
\newcommand\FFNLLmOne{\textcolor{blue}{\FFNLLmO{}\lessspace}}

\newcommand{\realOfCurly}[2]{\left\lbrace#1\right.\left\lmoustache#2\right\rbrace}
\newcommand{\mOneS}{M_{t}^{1S}}

\newcommand{\Op}{\mathcal{O}}
\newcommand{\na}{\textcolor{red}{N/A}}

\def\bq{\begin{equation}}
\def\eq{\end{equation}}
\def\ba{\begin{eqnarray}}
\def\ea{\end{eqnarray}}

\newcommand\Rcite[1]{ref.~\cite{#1}}
\def\refeq#1{\mbox{(\ref{#1})}}

\usepackage[babel]{csquotes}
\usepackage{pgf}
\usepackage{pgfkeys}
\expandafter\def\csname pgf@textdist@protect\endcsname{}
\input pgfmath.tex
\pgfset{
     number format/.cd,
     min exponent for 1000 sep=4,
     int detect,
  }

\newcommand{\kfactor}[2]{
  \def\x{#1}
  \def\y{#2}
  \def\xy{(\x) / (\y)}
  \pgfmathprint{\xy}
}

\newcommand{\tableline}[5]{
  \ifthenelse{\equal{#1}{}}
    {\na & \na & \na}
    {\ifthenelse{\equal{#3}{}}
      {
        \ifthenelse{\equal{#5}{}}
        {$#1 \pm #2$ & \na & \na}
        {$#1 \pm #2 \cdot 10^{-#5}$ & \na & \na}
            }
      {
        \ifthenelse{\equal{#5}{}}
        {$#1 \pm #2$ & $#3 \pm #4$ & \kfactor{#3}{#1}}
        {$#1 \pm #2 \cdot 10^{-#5}$ & $#3 \pm #4 \cdot 10^{-#5}$ & \kfactor{#3}{#1}}
      }
          }
}

\newlength{\jeroenlen}

\newcommand{\diagram}[1]{%
\begin{array}{l}
  \includegraphics[height=.13\textwidth]{Diags/#1-crop}
\end{array}
}

\begin{document}
\date{\today}

\titlehead{\hfill DESY 17-041}

\title{NLO QCD Corrections to Off-shell $t\,\bar{t}$ and $t\,\bar{t}\,H$ at the ILC}
\subtitle{\vspace{1em}Talk presented at the International Workshop on Future Linear Colliders
	  (LCWS2016), Morioka, Japan, 5-9 December 2016. C16-12-05.4.}

\author{J\"urgen Reuter\email{juergen.reuter}{desy.de},
        Bijan Chokouf\'e Nejad\email{bijan.chokoufe}{desy.de},
  Christian Weiss\email{christian.weiss}{desy.de}}

\publishers{\vspace{1em}\textit{
  DESY Theory Group,} \\
\textit{Notkestr. 85, D-22607 Hamburg, Germany}
\\[.5\baselineskip]}
\maketitle

\begin{abstract}
We discuss top-quark physics at the ILC with a focus on the full
  off-shell processes for $t\bar{t}$ and $t\bar{t}H$ production,
  including top-quark decays and also leptonic $W$ decays. A special
  focus is on the matching of the resummed vNRQCD threshold calculation
  and the fixed-order NLO QCD continuum calculation, where we present an
  update on the validation of the matching. All of the calculations have
  been performed in the \wz event generator framework.
\end{abstract}

\section{The \wz event generator at NLO}

\wz~\cite{Kilian:2007gr, Moretti:2001zz} is a multi-purpose event
generator for both lepton and hadron colliders. At leading-order, it
can compute arbitrary SM processes and supports a multitude of BSM
physics (e.g. using automated interfaces~\cite{Christensen:2010wz}).
For QCD processes, it uses the color-flow formalism \cite{Kilian:2012pz}. It has its
own implementation of an analytic parton shower \cite{Kilian:2011ka}.
Moreover, it can perform simulations for
a broad class of processes at next-to-leading order. The modern
release series (v2) has been developed to meet the demands of LHC
physics analysis, while its treatment of beam-spectra and
initial-state photon radiation makes it especially well suited for
lepton collider physics.

The generic next-to-leading order (NLO) framework in \wz builds upon the
FKS subtraction scheme~\cite{hep-ph/9512328, 0908.4272}, which
partitions the phase space into regions wherein only one divergent
configuration is present.
This divergence is then regulated using plus-distributions.
FKS subtraction synergizes with \wz{}'s optimized
multi-channel phase-space generator for the underlying Born
kinematics, from which real kinematics are generated. It is also
very well suited to the parton shower matching procedures employed, as
described below.
\wz{} supports \ol{}~\cite{Cascioli:2011va}, \go{}~\cite{Cullen:2014yla,Cullen:2011ac} and
\re{}~\cite{Actis:2012qn,Actis:2016mpe} as one-loop matrix element
providers as well as
for the computation of color- and spin-correlated Born matrix
elements. At tree-level, they can also be used as alternatives to
\wz{}'s standard matrix-element generator \om{} \cite{Moretti:2001zz}.

For event generation, \wz{} can produce weighted fixed-order NLO QCD
events that are written to e.g. \hepmc{}~\cite{Dobbs:2001ck} files. This
allows for flexible phenomenological fixed order studies, especially
in combination with \rivet{}'s~\cite{1003.0694} generic event analysis
capabilities. Matching to parton showers is achieved with an
independent implementation~\cite{1510.02739} of the \po{} matching
method~\cite{hep-ph/0409146}.

Apart from scattering processes, \wz is also able to compute decay
widths for \mbox{$1 \to N$} processes at NLO\@. The final-state phase space
is built in the usual fashion, whereas the initial-state phase space
is adapted for decays. Computing decay widths directly in \wz
allows for a consistent treatment of top and gauge boson widths in
an NLO calculation.

\section{The \ttb and \tth continuum at NLO QCD}

The new \wz FKS implementation has been applied to an extensive study of
fully off-shell $\ttb$ and $\tth$ production at a lepton collider~\cite{Nejad:2016bci}.
Top-quark and leptonic $W$ decays are taken into account including the
full irreducible background. The (loop) matrix elements are obtained
from \ol, which has been applied to a lepton collider process
including hexagon diagrams for the first time.
Moreover, the resonance-aware modification of FKS subtraction~\cite{Jezo:2015aia} is used
to treat intermediate top, Higgs and $Z$ resonances.

\begin{figure}[htbp]
\centering
\includegraphics[width=0.45\textwidth]{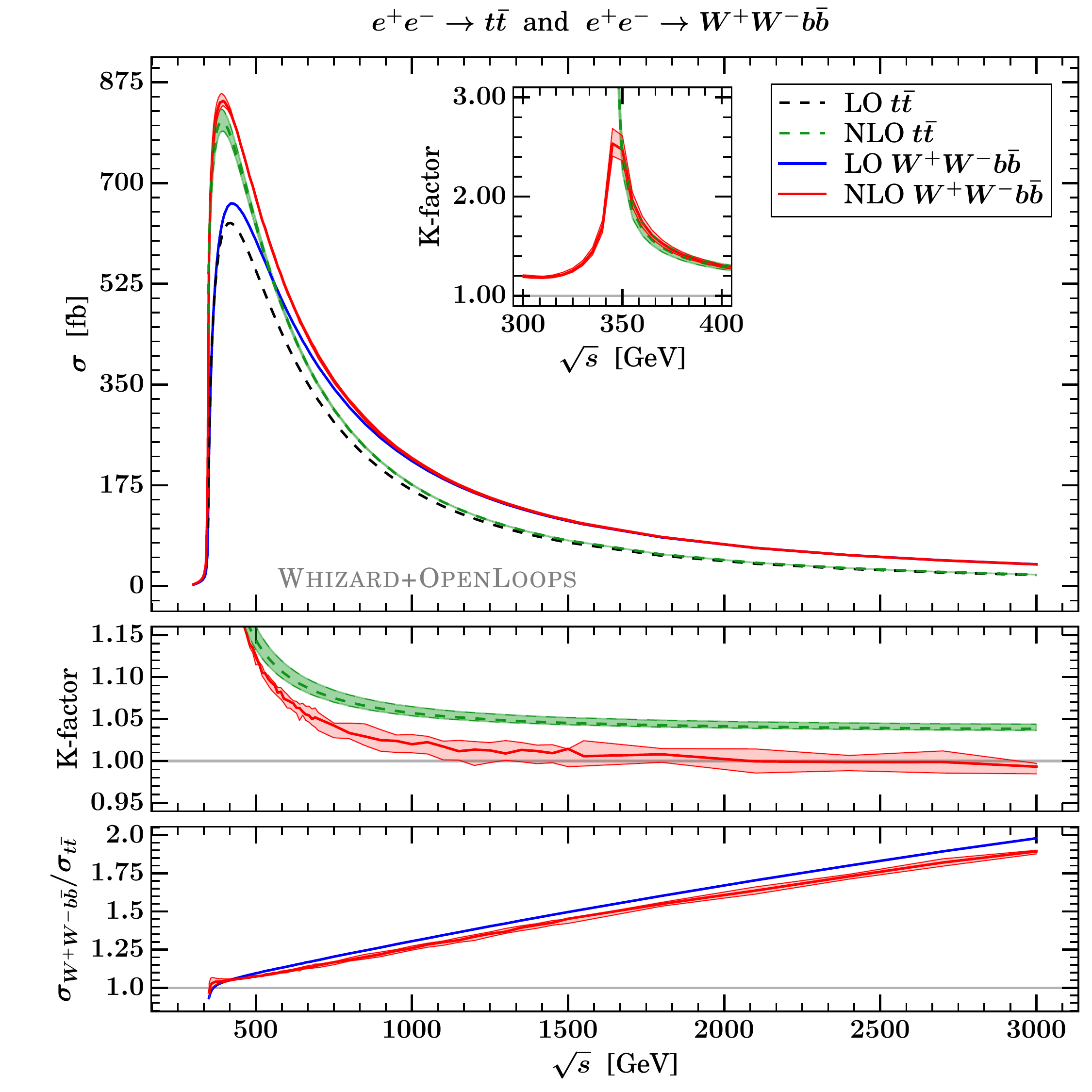}
\includegraphics[width=0.45\textwidth]{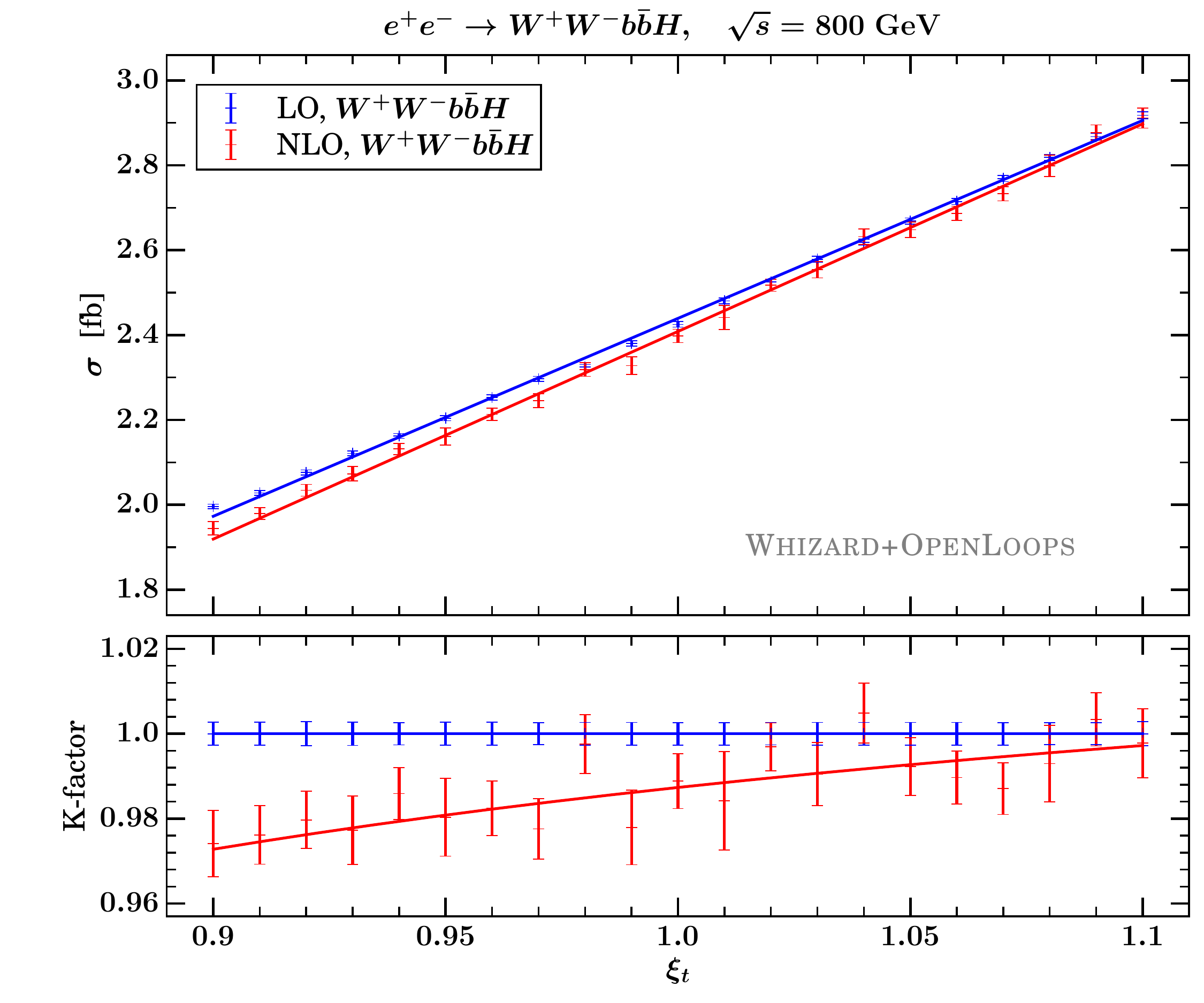}
\caption{In the left plot, we show the total cross section for on-shell
  and off-shell $t\,\bar{t}$ production  as a function of $\sqrts$. In the
  lower panels, we display the K-factor for $t\,\bar t$ and $\wbwb$ in
  green and red, respectively, as well as the ratio of off-shell to
  on-shell results for LO and NLO in blue and red.
  In the right plot, we present the \eewwbbH{} LO and NLO cross sections
  as a function of the top Yukawa coupling modifier $\xi_t=y_t/y_t^\text{SM}$, as well as a linear fit.
  }
\label{fig:continuum}
\end{figure}

On the left-hand side of \cref{fig:continuum}, we show a scan of the
total inclusive cross section for the on-shell
process $\epem \to \ttb$ and the off-shell process $\epem \to \wbwb$ as computed by \wz.
The most striking feature is that right above the production threshold
$\sqrts = 2m_t$, both LO and NLO cross sections are strongly enhanced. Moreover, in
the limit $\sqrts \to 2m_t$ the NLO corrections to the on-shell process diverge
due to non-relativistic threshold corrections, which manifest themselves as large
logarithmic contributions to the virtual one-loop matrix element. In the off-shell
process, the Coulomb singularity is regularized by the top-quark width, so that
NLO corrections remain finite. Nevertheless, threshold corrections introduce a distinct peak
in the K-factor at $\sqrts = 2m_t$, with a maximimum of about $2.5$.

The process $\epem \to \tth$ provides a unique opportunity to measure the
top Yukawa coupling $y_t$ \cite{Agashe:2013hma,1409.7157} at the per cent level. Many new
physics models, such as
generic 2HDMs, the MSSM or composite and Little Higgs models, predict significant
deviations of $y_t$ from its standard model value $y_t^\mathrm{SM} = \sqrt{2}m_t/v$.
The right-hand side of \cref{fig:continuum} shows the dependence
of the off-shell process on $y_t$, parametrized as $y_t=\xi_t
y_t^\mathrm{SM}$, both at leading and next-to-leading order. The linear
fit can be used to extract the parameter $\kappa$, defined via \cite{1409.7157,1307.7644}
\begin{equation}
\label{eq:yukawa-dep}
  \frac{\Delta y_t}{y_t} = \kappa \frac{\Delta \sigma}{\sigma}\;.
\end{equation}
$\kappa$ contains contributions from signal, background and inteference terms. Since
the $y_t$-dependence of the cross section on $y_t$ is approximately quadratic, $\kappa$ is
close to $0.5$. In the above plot, we find NLO QCD corrections to $\kappa$ to be
significant. They decrease $\kappa$ from the value $0.52$ at LO by about $4.6\%$ to
$\kappa = 0.497$ at NLO. A detailed analysis~\cite{Nejad:2016bci} reveals that these negative
corrections have to originate from interference terms.

Far above the threshold, the NLO corrections are rather small for both the on-shell
and the off-shell processes. For \eett, the corrections remain positive for all \sqrts,
approaching the universal massless quark pair-production factor $\alpha_s / \pi$ as
the top mass becomes negligible. In contrast, the NLO corrections to \eewwbb decrease
significantly faster for large \sqrts, are at the per cent level for $\sqrts=\ValGeV{1500}$,
and come close to zero at $\sqrts=\ValGeV{3000}$.

\section{Top-Quark threshold resummation and NLO matching}

The large NLO corrections encountered in the previous section are well-known
to arise from gluon exchange in the virtual correction
to the top-quark production diagram. They appear as logarithms of the non-relativistic
velocity $v$ and the strong coupling $\alpha_s$, which can be resummed.
One approach for this is
vNRQCD~\cite{hep-ph/9910209,hep-ph/9912226,hep-ph/0003032,hep-ph/0003107},
in which an effective Lagrangian for the interaction of
non-relativistic heavy quark pairs is constructed. The result of the
resummation can, up to NLL, be included as a simple form factor $F_i$ for
$t\,\bar{t}$ production.
Hereby, $i = \{\mathrm{LL},\mathrm{NLL}\}$ denotes the order of resummation.
The vNRQCD results can be used in \wz{} by embedding $F_i$ within a
gauge-invariant description of $t\,\bar{t}$ production, as elaborated
further below.
In this section, we report on the recent development of the combination
of the resummation with fixed-order NLO results to achieve a consistent
treatment of top production at a lepton collider at all center-of-mass
energies.

\subsection{Setup of the calculation}

\subsubsection{Relativistic embedding of the form factor}
\begin{figure}[htbp]
\centering
\includegraphics[width=0.45\textwidth]{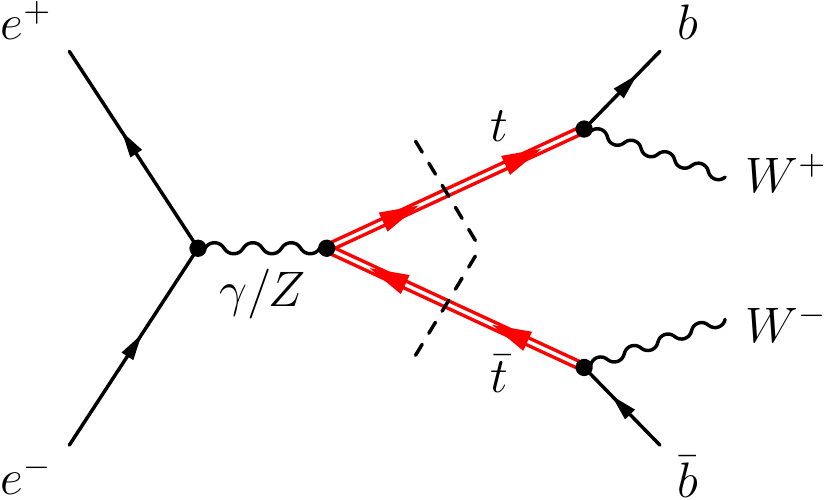}
\caption{Depiction of the factorized computation in the double pole
  approximation. Double lines indicate top propagators and a dashed
  line crossing them a factorized computation with on-shell projection.}
\label{fig:factorized}
\end{figure}

The resummed form factor is included in a gauge invariant way by
factorizing the full matrix element into a production and a decay
contribution,
\begin{equation}
\label{eq:factorized}
\ME =
   \underbrace{\braketop{\epem}{\mathcal{T}_\T{NRQCD}}{{\ttb}}}_{\equiv \MEprod}
   \braketop{{\ttb}}{\mathcal{T}}{\wbwb},
\end{equation}
where the form factor only enters the production matrix element \MEprod.
The remaining factor $\braketop{{\ttb}}{\mathcal{T}}{\wbwb}$ contains
propagators and decay matrix elements for both top-quark lines.
\Cref{eq:factorized} is represented diagrammatically in \cref{fig:factorized}.
Specifically, we are using a double-pole
approximation (DPA)~\cite{Stuart1991,hep-ph/9312212,hep-ph/9811481,hep-ph/9912261}.
Hereby, the momenta of the top quarks and their decay products have to
be projected on-shell in the matrix elements to remove gauge-dependent
contributions.
In the denominators of the top propagators and the phase-space
Jacobians, the off-shell momenta are used.
We extend the DPA also below threshold by evaluating the matrix elements
with momenta at threshold.
This can be seen as the closest gauge-invariant extension of the DPA
that is non-zero below threshold.
For comparison, we also show results in the validation that can be
obtained if a gauge-dependent approach, i.e.\ signal diagram with
off-shell momenta, is used.

\subsubsection{Matching}

The matching procedure combines the (N)LL expressions $\sigma_{\mathrm{NRQCD}}$,
including the (N)LO decay, with the full fixed-order (N)LO results $\sigma_{\mathrm{FO}}$ for
\wbwb including all irreducible background processes and interferences.

By construction, the resummed result is only a valid approximation for
$v \sim \alpha_s$.
Its contribution, therefore, has to become negligible for
$\abs{\sqrts-2m_t}\gg \Gamma_t$.
This can be achieved by introducing a switch-off function $f_s(v)$,
which is multiplied to each strong coupling constant in the resummed
computation\footnote{
	$\switch$ can in principle also be directly multiplied to the matrix elements,
	yet associating them with the couplings ensures a smoother switch-off.
}.
The explicit form is arbitrary, with the minimal requirements
\begin{equation}
\label{eq:switch-off}
  f_s\left(v_\T{min}\right) = 1 \quad \text{and} \quad f_s(v=1) = 0,
\end{equation}
whereby the velocity $v$ takes its minimal value at threshold. Due to the
presence of the width, $v_\T{min} \sim 0.1 > 0$.
For a realistic
phenomenological description, we will switch off not too close to threshold
in order to use resummed results in a region as wide as possible, but also
not too far from threshold where any NRQCD loses predictivity and validity.
The next cornerstone of the matching procedure is the treatment of the
first order in $\alpha_s$.
As both the resummed and the fixed-order result contain it, a naive
addition of both results yields a double counting of
$\Op(\alpha_s)$-terms.
To solve this problem, we use \XSexpanded, the resummed cross
section expanded to $\Op(\alpha_s)$.
Thus, the master formula for the matched cross section is
\begin{align}
  \label{eq:matched_simple}
  \XSmatched = \XSfo\of{\AShard}
  &+ \XSresummed\of{\switch\,\AShard,\;\switch\,\ASsoft,\;\switch\,\ASusoft} \no
  &- \XSexpanded\of{\switch\,\AShard},
\end{align}
where in the full NRQCD calculation, the strong coupling has to be
evaluated at hard (H), soft (S) and ultra-soft (US) scales.
To remove the double counting and to ensure
the NLL validity of \XSmatched, \XSexpanded has to be evaluated at the
same (hard) scale as \XSfo. Note that in
eq.~\refeq{eq:matched_simple}, all strong couplings in the NRQCD terms are
already multiplied with $f_s$.

Diagramatically, eq.~\refeq{eq:matched_simple} takes the form
\begin{align}
\label{eq:NLONLL}
  \XSNLONLL &= \XSNLO + \realOfDiagrams[\textcolor{blue}
      {\left(\FFNLLmO - \FFexpNLLmOne\right)}\hspace{-1.3em}]
      {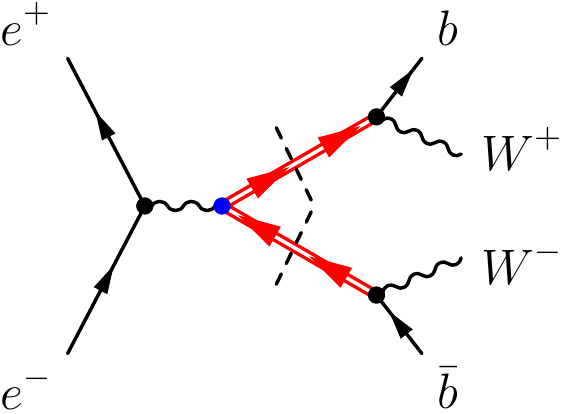}{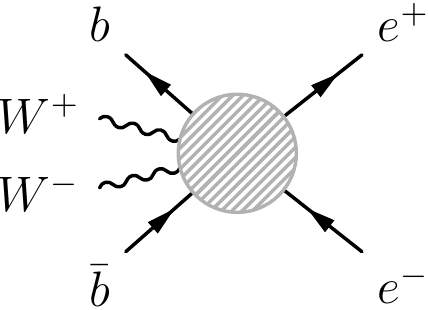}\no
      &+ \abs{\FFNLLmOne\diagram{factorizedLessDetail}}^2 \no
      &+ \realOfCurly{\FFNLLmOne{}\extraspace{}\left(\diagram{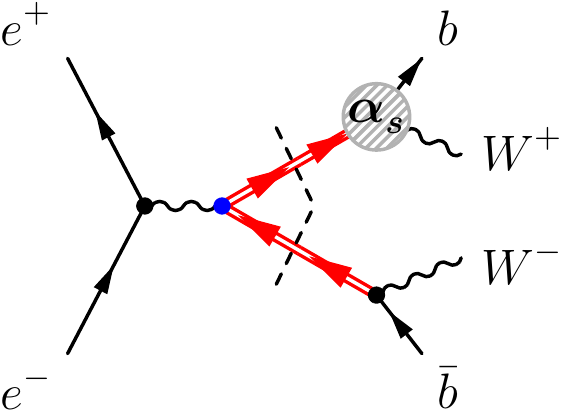}
                        + \lessspace\diagram{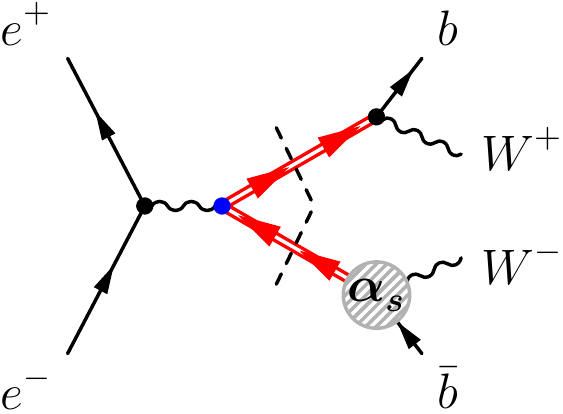}\right)}
        {\diagram{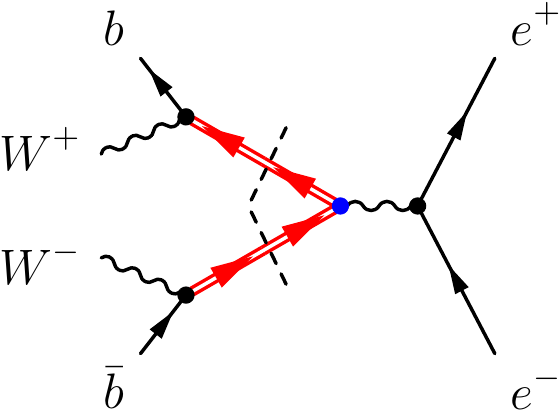}\lessspace{}\FFNLLmOne{}\extraspace{}}\no
      &+ \abs{\FFNLLmOne{}\diagram{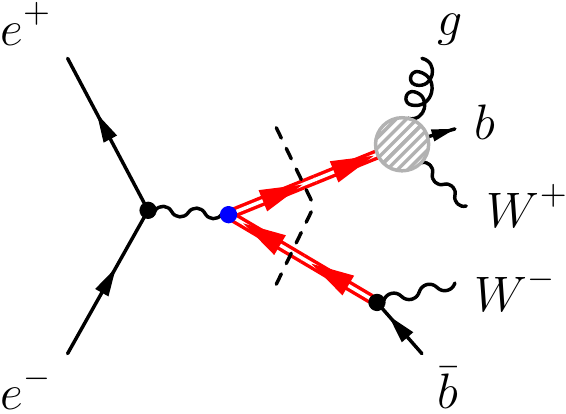}}^2
       + \abs{\FFNLLmOne{}\diagram{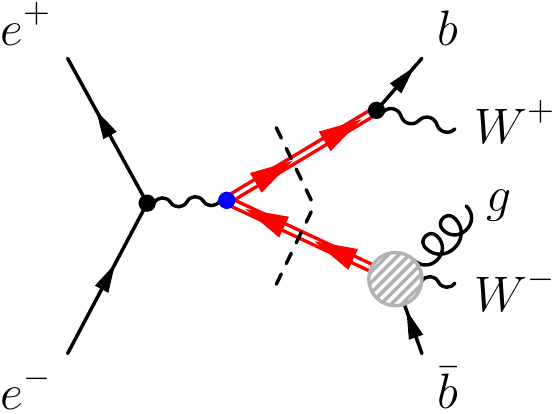}}^2,
\end{align}
with $\FFNLLmO{} = \FFNLL - 1$.
The first summand after \XSNLO is the interference term between the
factorized computation,
\cref{eq:factorized}, and the full LO amplitude, including
all $2 \to 4$ contributions, indicated by the gray blob.
This term contains both the full form factor as well as its expansion.
On the second line, we have the square of $\tilde F$, which is followed
by its hard NLO corrections to the top decay:
In the third line, we find the virtual component, indicated by the small
gray blob with $\alpha_s$ inside.
They operate only on the legs they are attached to, i.e. each blob
consists of one gluon loop connecting the bottom and top quark.
Finally, in the last line, there are the squared real amplitudes.
Here, each gray blob represents two diagrams for gluon emission from the
bottom and the top quark, respectively.
Note that interference terms between the real diagrams are discarded, as
they would introduce infrared divergences not cancelled by the virtual
diagrams.

\subsection{Implementation in \wz}
The form factor only has an analytical expression at LL, while at higher
logarithmic orders, only numerical computations are possible. A dedicated
tool for this is \toppik~\cite{hep-ph/9904468},
which is included in the \wz distribution.

The factorized tree-level matrix elements are calculated by modified
\om codes. We obtain loop matrix elements from \ol.
For this purpose, a dedicated matrix element library for polarized top
decays is used, especially for spin correlations in top decays, which is publicly available.
The one-loop decay amplitudes are then combined with the same code for
the production matrix element as for tree level amplitudes.

We use the FKS setup of \wz to evaluate eq.~\refeq{eq:NLONLL}.
The treatment of the fixed-order NLO cross section \XSNLO is identical
to the previous section and \Rcite{Nejad:2016bci}. Thus, we can
use the standard algorithm and add the result to the rest of the
formula.
For the remainder, slight modifications have to be made to the
subtraction.
They can be summed up as the following.

\paragraph{On-shell generation of the real-emission phase space}
Like the Born matrix element, the real-emission matrix element has to be evaluated
using on-shell momenta. In FKS, the phase space with an additional gluon $\Phi_{n+1}$ is constructed
based on the underlying Born phase space for each possible emitter. Therefore,
we already start with an on-shell projected phase space. The emission mapping
then has to ensure that this property is kept. For this purpose, we use the same
phase-space construction as in the resonance-aware FKS approach. There, $\Phi_{n+1}$
is constructed so that the invariant mass of the resonance associated with the
emission is conserved. Fixing the invariant mass automatically ensures that an
on-shell phase space stays an on-shell phase space, so that we just adopt the
same mappings outlined in \Rcite{Jezo:2015aia}.
\paragraph{Decay subtraction}
The divergences in the factorized calculation all originate from the $t \to bWg$ matrix element.
It consists of two Feynman diagrams, one in which the gluon is emitted from the top
quark and another one in which it is emitted from the bottom. Divergences can
only occur in emissions from particles with on-shell momenta and zero width. Therefore,
in the full \wbwb matrix element, emissions from internal top quarks do not yield
divergences, as they are regulated by the width. However, in the factorized approach,
the gluon emission from the top quark is a singular contribution, which needs to
be subtracted. We call this additional singular region a pseudo-ISR region.
This way, each FKS pair $(b,g)$ and $(\bar{b},g)$ is associated with a pseudo-ISR pair
$(b,g)^*$ and $(\bar{b},g)^*$, in which the gluon radiation occurs not from the bottom,
but from the top quark. This means that in the corresponding singular region, the FKS
phase-space contribution $d_{ij}$ is evaluated with $p_i \to p_{\mathrm{top}} = p_b + p_W$.
\paragraph{Omission of interference terms}
As outlined above, interference terms between emissions from different top-quark lines
are not included in our calculation. Therefore, they also need to be dropped from
the soft expressions in which mixed-emitter eikonal integrals appear.

\subsection{Validation and results}
\begin{figure}[htbp]
\centering
\includegraphics[width=0.43\textwidth]{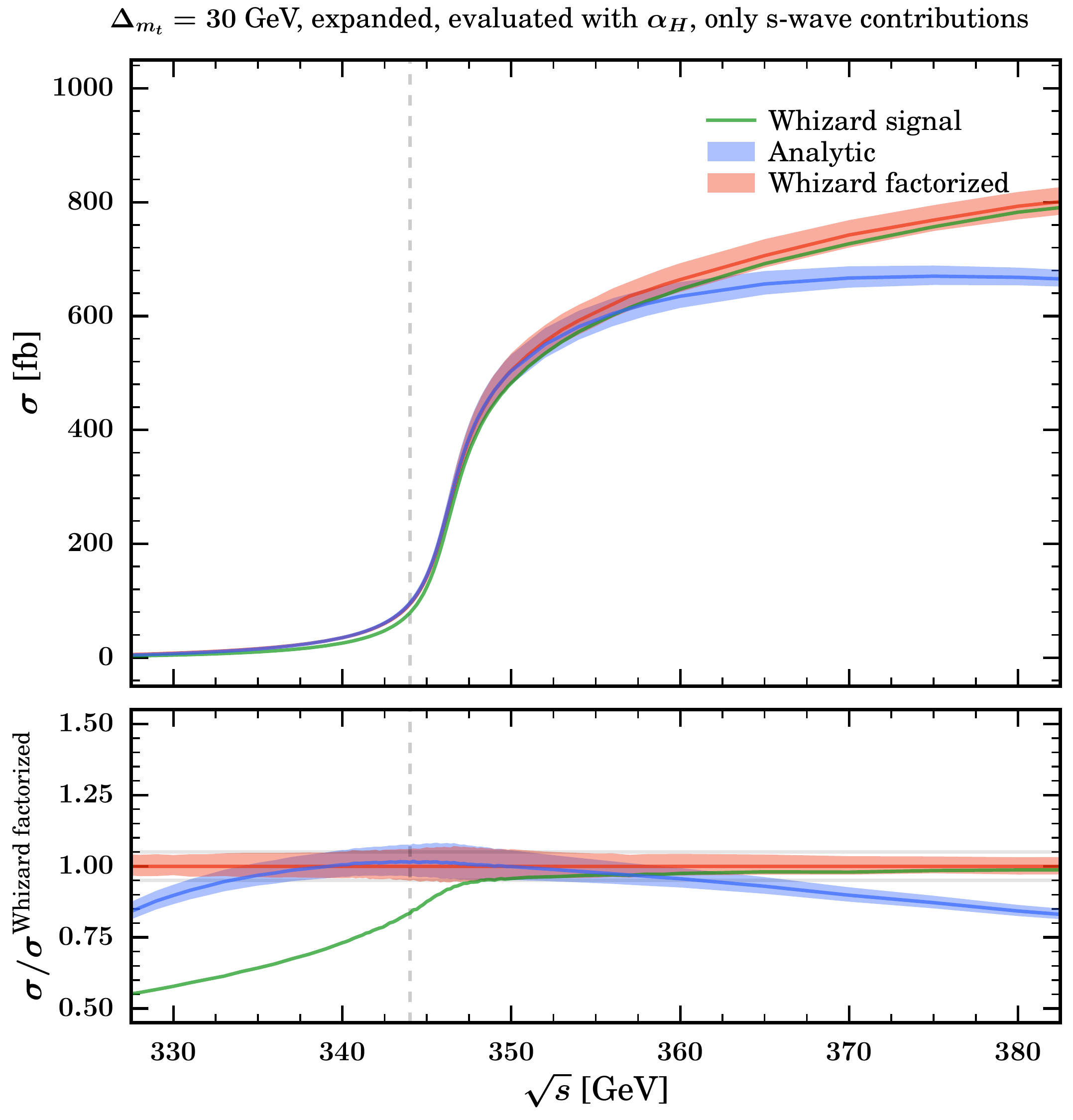}
\includegraphics[width=0.43\textwidth]{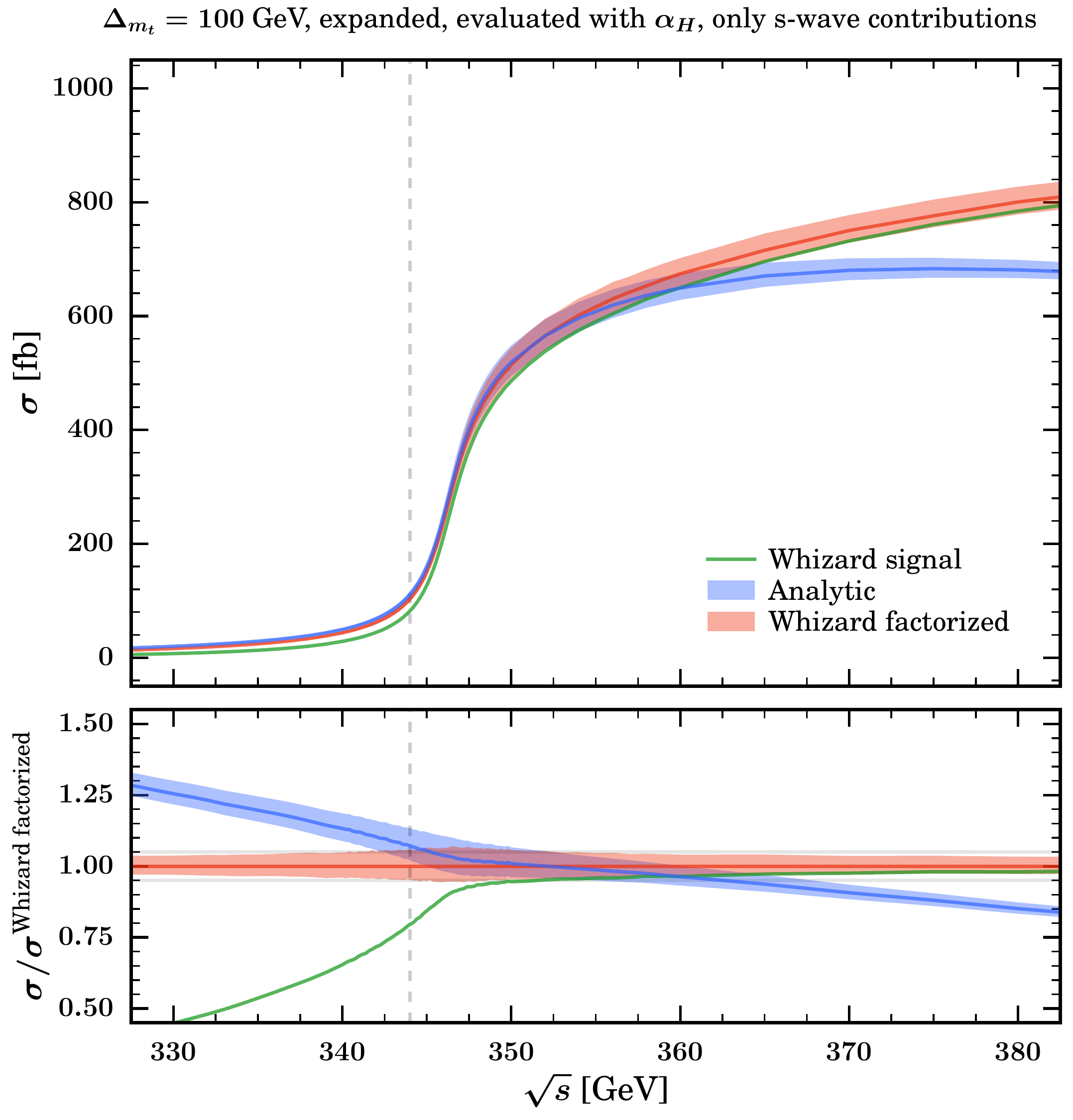}\\
\includegraphics[width=0.43\textwidth]{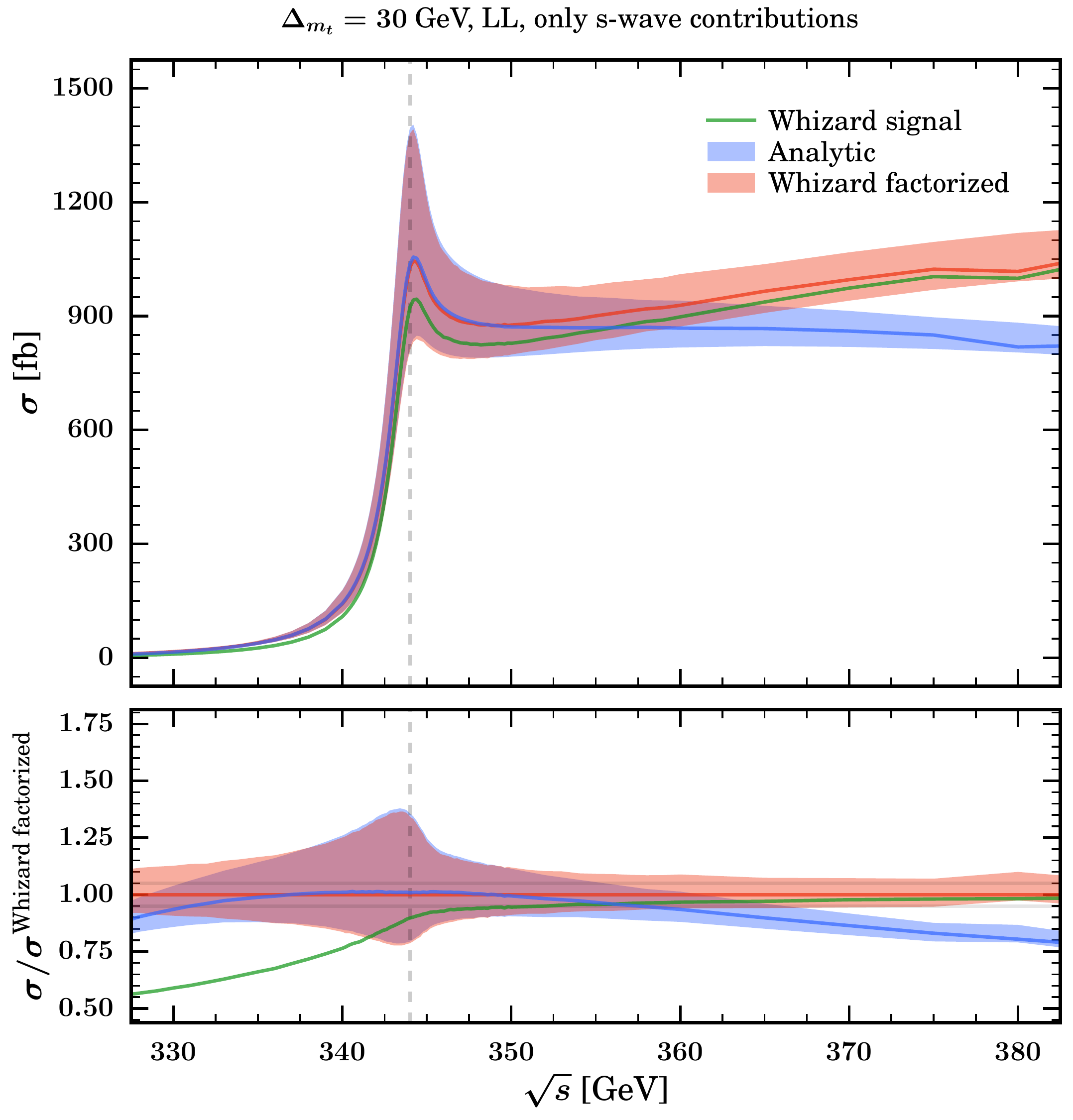}
\includegraphics[width=0.43\textwidth]{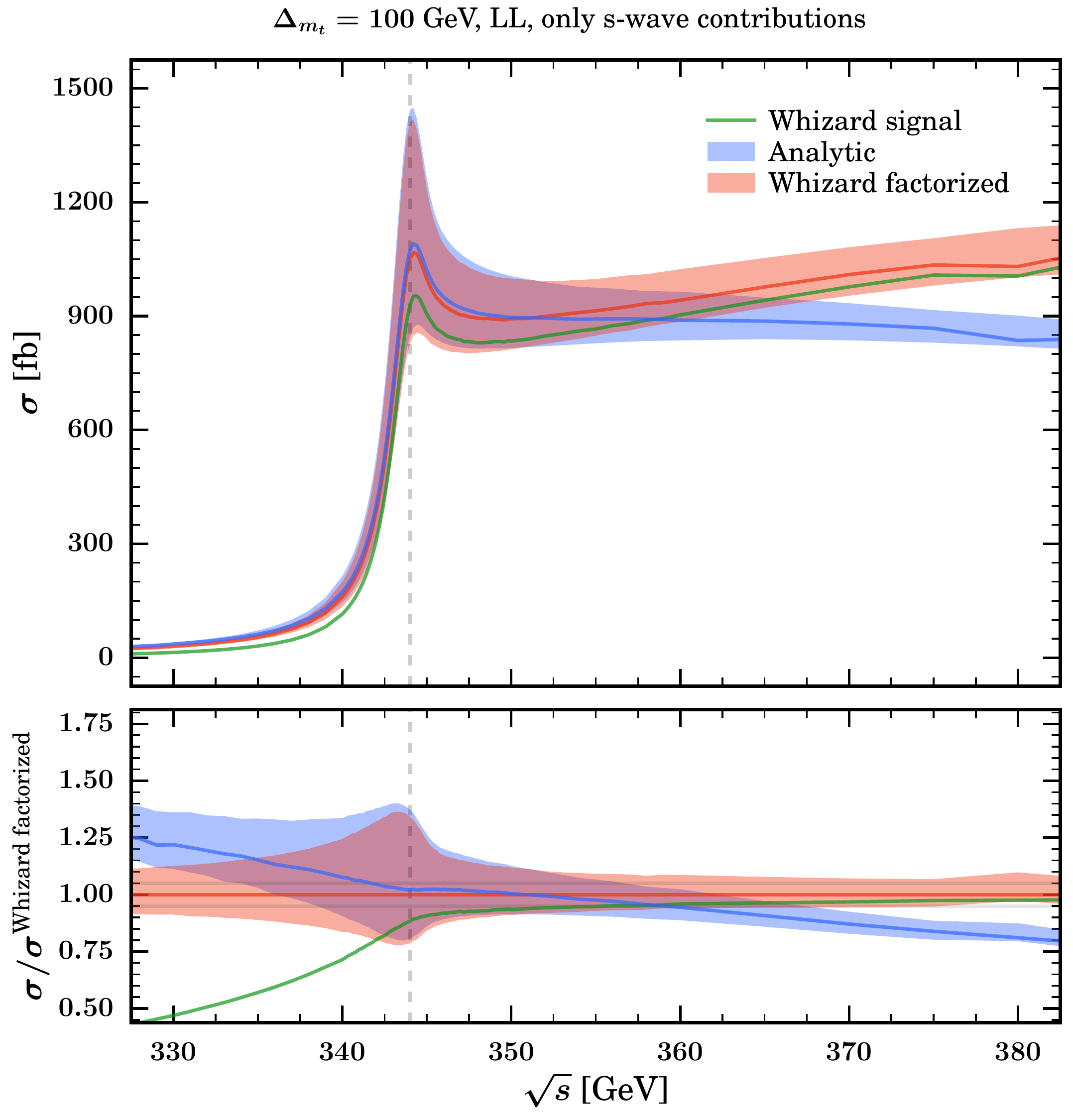}\\
\includegraphics[width=0.43\textwidth]{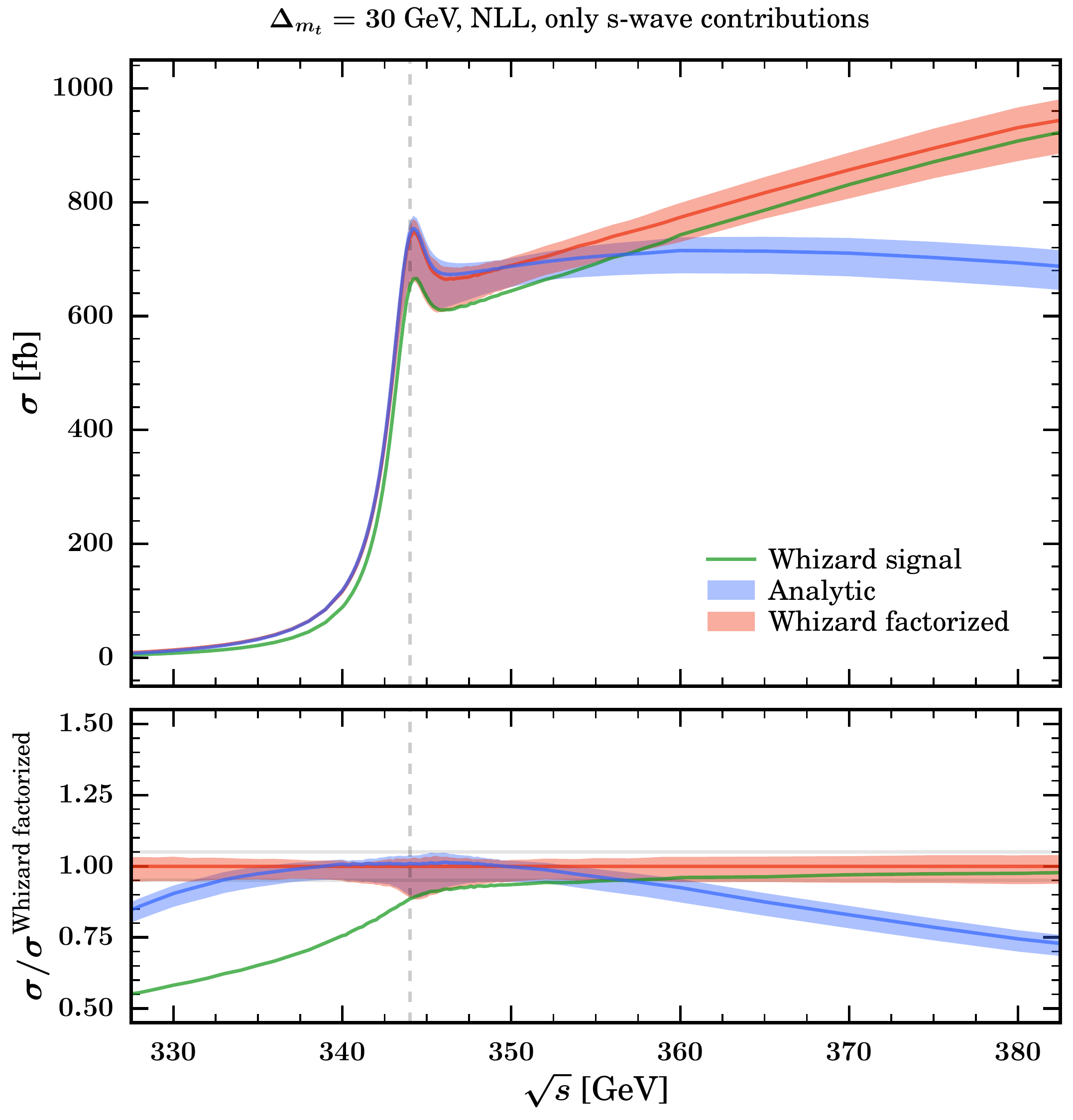}
\includegraphics[width=0.43\textwidth]{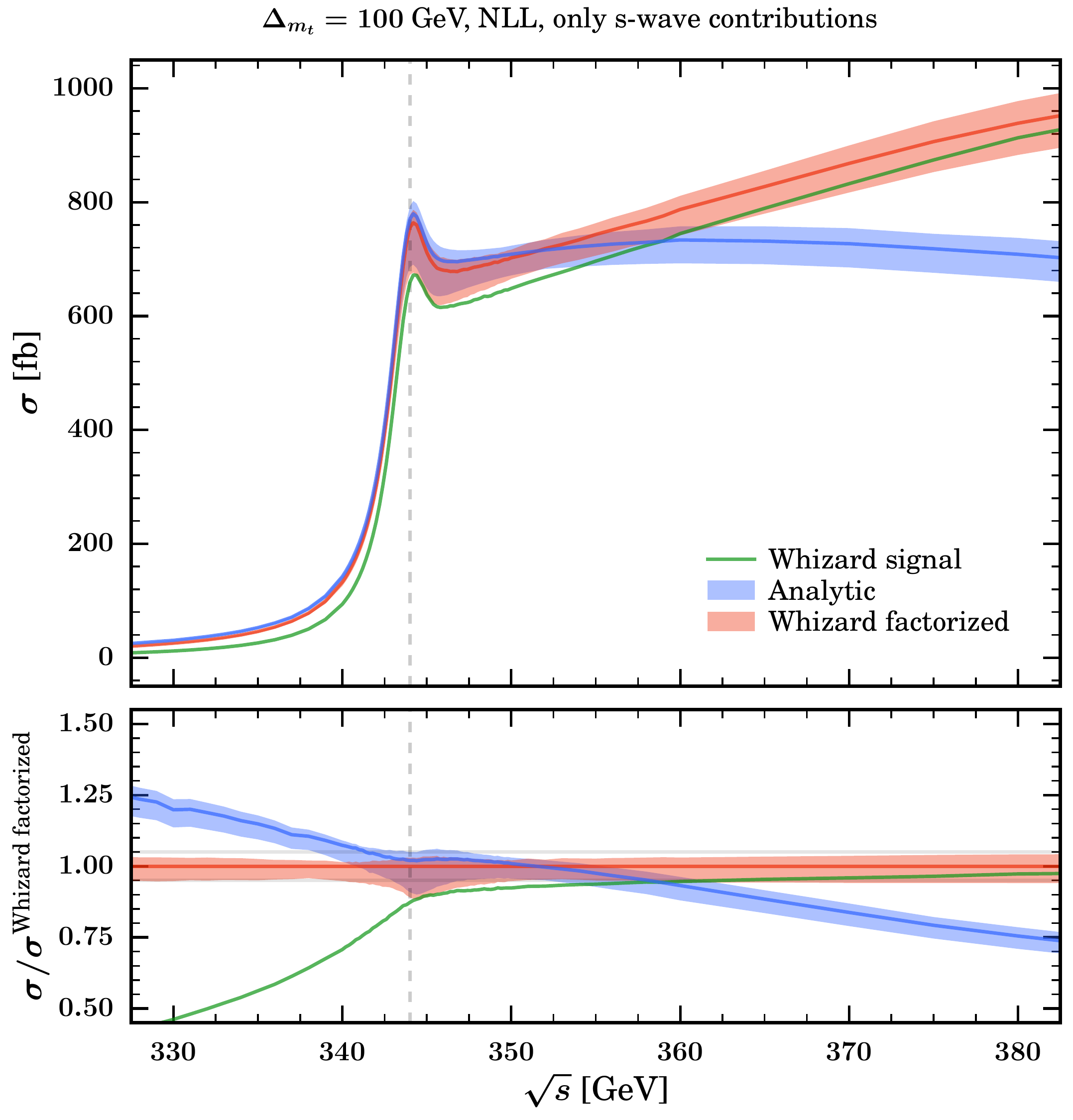}
\caption{Comparison of analytic results with the implementation in \wz with
  the factorized and the signal-diagram approach for $\Delta_{m_t} =
  \ValGeV{30}$ and $\Delta_{m_t} = \ValGeV{100}$ using an expanded, LL or
  NLL form factor.
  The bands correspond to the envelope of the scale variations mentioned
  in the text.
  }
\label{fig:validation_dm_fixed}
\end{figure}
The implementation in \wz can be checked against the analytical
calculation of \Rcite{1309.6323}.
For reliable numerical predictions, a cut $\Delta_{m_t}$ on the reconstructed top
invariant mass is required~\cite{1002.3223}, fulfilling
\begin{equation}
\label{eq:DeltaMcut}
  \left|\sqrt{\left(p_{W^+} + p_b\right)^2} - \mOneS\right| \leq \Delta_{m_t} \quad \text{and} \quad
  \left|\sqrt{\left(p_{W^-} + p_{\bar{b}}\right)^2} - \mOneS\right| \leq \Delta_{m_t}\;.
\end{equation}
We stress that although this cut depends on $\mOneS$, the invariant mass
distributions will be centered around the pole mass $m_t$. While
\cref{eq:DeltaMcut} is exact in \wz, in the analytic calculation, we
implement a cut on the nonrelativistic invariant masses,
\begin{equation}
\label{eq:def-nonrel-mass}
  t_{1,2} = 2m_t \left(E_{1,2} - \frac{\vec{p}\,^2}{2m_t}\right),
\end{equation}
by requiring that
\begin{equation}
\label{eq:cut-def-nonrel}
  \left|t_{1,2}\right| \leq 2\mOneS \Delta_{m_t} - \frac{3}{4}\Delta_{m_t}^2 + \Op(v^2).
\end{equation}
Here, $E_{1,2}$ are the kinetic energies of the top and anti top quark, respectively,
and $\vec{p}$ is the top quark three momentum. These different cut implementations are
one source of disagreement between the Monte Carlo and the analytic results. In the
threshold region, the difference should, however, be of higher order.

\begin{figure}[htbp]
\centering
\includegraphics[width=.8\textwidth]{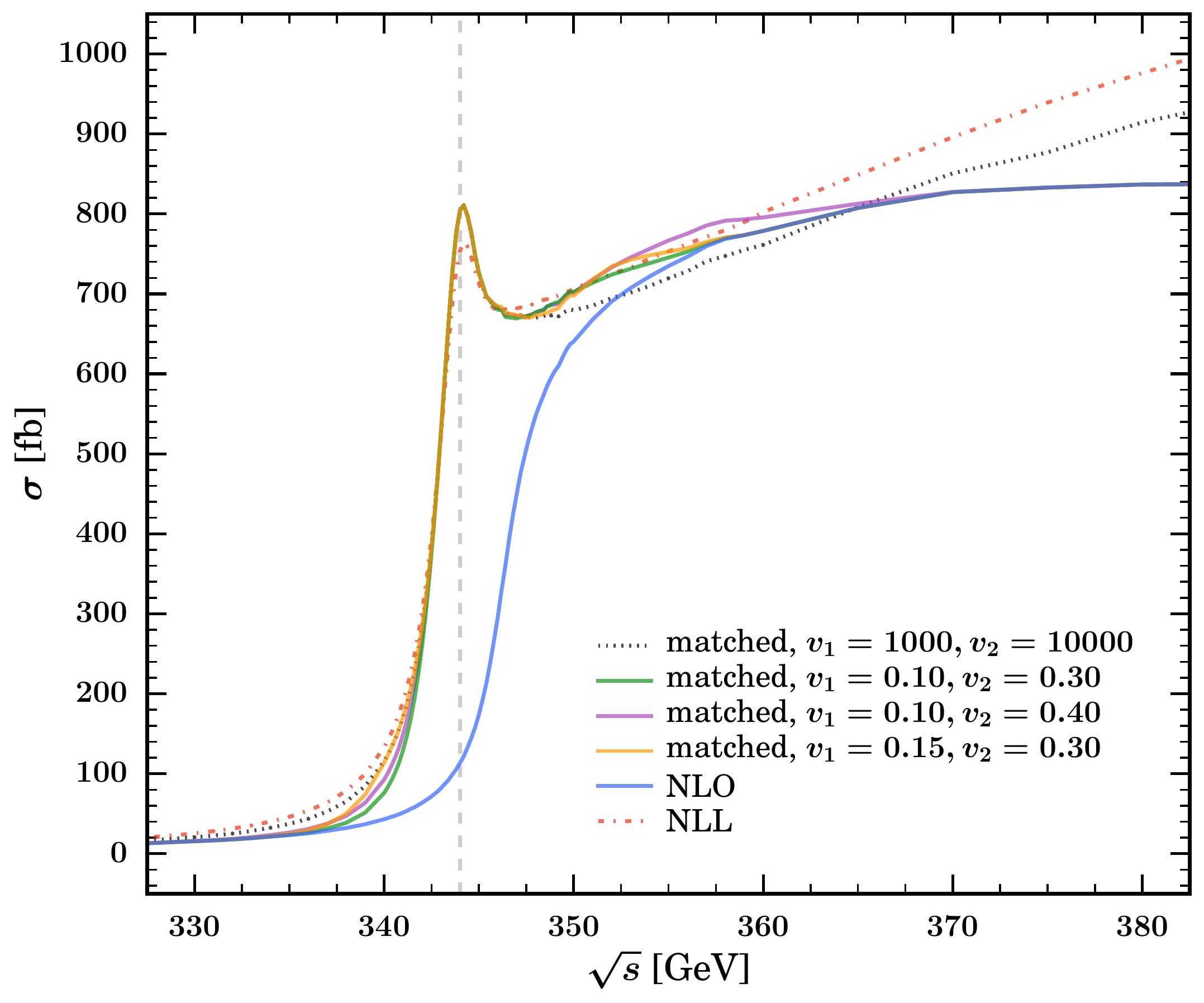}
\caption{The fully matched total cross section for $\epem \to \wbwb$
  including NLO decays, the NLL form factor and the full NLO computation
  according to \cref{eq:NLONLL}.
  In addition to the three curves that are obtained for
  each of the three choices of the matching parameters $(v_1,v_2)$, we show the curve of pure
  fixed-order NLO and lines for NLL (red, dashed) and the matched result
  without switching off (black, dotted).
  }
\label{fig:matched}
\end{figure}

In \cref{fig:validation_dm_fixed}, we show \sqrts-scans for a
fixed value of $\Delta_{m_t}$. 
We have two different cut choices, a moderate, $\Delta_{m_t} =
\ValGeV{30}$, and a loose cut, $\Delta_{m_t} = \ValGeV{100}$.
A detailed analysis shows that the analytic computation is only reliable
for moderate cuts.
The plots in fig.~\ref{fig:validation_dm_fixed}, show perfect agreement
between the analytic computation and \wz for the moderate cut
(${\Delta_{m_t} = \ValGeV{30}}$) within a window around threshold of at
least $\ValGeV{10}$. For the loose cut, this range is reduced due to
additional nonphysical contributions below threshold in the analytic
results.
For comparison, we also show the gauge-dependent results that can be
obtained when embedding the form factor naively into the signal diagram,
which leads to systematically lower results.

Finally, in fig.~\ref{fig:matched}, we present the matched total cross
section as a scan over \sqrts around threshold.
The matched curve is similar to the pure NLL computation with LO decay
around $\sqrts = 2\mOneS{}$ and then smoothly approaches the fixed-order
line.
To estimate the error due to the arbitrary switch-off function, we have
performed the computation for different values of start, $v_1$, and end,
$v_2$, of the switch-off.
We have experienced, furthermore, that shifting the switch-off
parameters to significantly lower values, like $v_1=0.1, v_2=0.2$, cuts
away too much of the threshold region and is far from the matching curve.
Note that we have used $\mOneS\sqrt{\abs{v}}$ as hard scale for $\XSfo$
and \XSexpanded in eq.~\refeq{eq:matched_simple}
instead of the more conventional hard scale $\mOneS$.
This is the geometric mean of the hard and the soft scale and thus a
more consistent choice if one aims to combine NLL and NLO results.
With this choice NLO and NLL approach each other and overlap at
$\sim\SI{357}{\GeV}$.
After this overlap, we expect the NLO to give the more reliable results
for higher $\sqrt{s}$.
Thus, higher values of the switch-off parameters are in principle
possible but likely unnecessary.
Overall, we observe fairly mild matching variation uncertainties as long
as it contains the important physical regions.
Finally, we want to emphasize that the matched computation, even without
switch-off, realized as ($v_1=1000,v_2=10000$), does not have to be in between NLL
and NLO as it is not a naive interpolation of these results but the
implementation of \cref{eq:NLONLL}.


\end{document}